\documentclass[preprint,journal]{vgtc}       

\ifpdf
  \pdfoutput=1\relax                   
  \usepackage{graphicx}                
  \DeclareGraphicsExtensions{.pdf,.png,.jpg,.jpeg} 
\else
  \usepackage{graphicx}                
  \DeclareGraphicsExtensions{.eps}     
\fi%

\graphicspath{{figures/}{pictures/}{images/}{./}} 

\usepackage{microtype}                 
\PassOptionsToPackage{warn}{textcomp}  
\usepackage{textcomp}                  
\usepackage{mathptmx}                  
\usepackage{times}                     
\usepackage{cite}                      
\usepackage{tabu}                      
\usepackage{booktabs}                  
\usepackage{enumitem}
\usepackage{booktabs} 
\usepackage{makecell}
\usepackage{tabularx}
\usepackage{upgreek}
\usepackage{wrapfig}
\usepackage{float}
\usepackage{subfloat}
\usepackage[bookmarks=false]{hyperref}

\onlineid{0}

\vgtccategory{Research}
\vgtcpapertype{please specify}

\title{Characterizing the Quality of Insight by Interactions: A Case Study}

\author{Chen He, Luana Micallef, Liye He, Gopal Peddinti, Tero Aittokallio, and Giulio Jacucci}
\authorfooter{
\item
 Chen He and Giulio Jacucci are with the Department of Computer Science, University of Helsinki. E-mail: chen.he@helsinki.fi and giulio.jacucci@helsinki.fi.
\item
Luana Micallef was with Human-Centered Computing, Department of Computer Science, University of Copenhagen.
\item
Liye He and Tero Aittokallio are with the Institute for Molecular Medicine Finland, University of Helsinki. E-mail: liye.he@helsinki.fi and tero.aittokallio@helsinki.fi.
\item
 Gopal Peddinti is with the VTT Technical Research Center of Finland Oy. E-mail: gopal.peddinti@vtt.fi.
}

\shortauthortitle{He \MakeLowercase{\textit{et al.}}: Characterizing the Quality of Insight by Interactions: A Case Study}

\abstract{Understanding the quality of insight has become increasingly important with the trend of allowing users to post comments during visual exploration, yet approaches for qualifying insight are rare. This paper presents a case study to investigate the possibility of characterizing the quality of insight via the interactions performed. To do this, we devised the interaction of a visualization tool---MediSyn---for insight generation. MediSyn supports five types of interactions: selecting, connecting, elaborating, exploring, and sharing. We evaluated MediSyn with 14 participants by allowing them to freely explore the data and generate insights. We then extracted seven interaction patterns from their interaction logs and correlated the patterns to four aspects of insight quality.
The results show the possibility of qualifying insights via interactions. Among other findings, exploration actions can lead to unexpected insights; the drill-down pattern tends to increase the domain values of insights. A qualitative analysis shows that using domain knowledge to guide exploration can positively affect the domain value of derived insights. We discuss the study's implications, lessons learned, and future research opportunities.}


\vgtcinsertpkg

\begin{document}


\firstsection{Introduction}

\maketitle

Insight in visualization denote an advance in knowledge or a piece of information \cite{defining}. Gaining insight is considered one of the major purposes of visual data exploration \cite{understanding}. 
Hence, understanding the user insight generation process when using visualization tools is an important goal of visualization research \cite{logSurvey, provenance}. For instance, Yi et al. \cite{understanding} identified four categories of insight gaining processes through an extensive literature review: providing an overview, adjusting the level of abstraction and/or the range of selection, detecting patterns, and matching the user's mental model of the data. Visualization can support these processes to foster insight \cite{promoting}.

As insight generation is usually aided by user interaction in visualization, Reda et al. \cite{modeling} and Guo et al. \cite{casestudy} explored which types of interactions foster or hinder insight via interaction logs and verbal transcripts. In this paper, instead of investigating the relations between interaction types and the \textit{quantity} of insight, we seek to characterize the \textit{quality} of insight by interactions. 

The trend of allowing users to post comments during visual exploration (e.g., \cite{manyeyes, voyager, insideinsights})
makes the quality of generated insights more critical than the quantity. Wang et al. \cite{knowledge} suggested that unrelated or incorrect insights would degrade the benefits of representing insights in a visualization. However, verifying and validating diverse insights is challenging by nature \cite{knowledge}. We suggest that the quality of insight can be relevant to the types of interactions performed. The interactions performed result from user intent \cite{mental, analyticprove}, whereas user intent may further affect the quality of insight. For instance, with the intent of taking an in-depth look at an issue, a user may retrieve the details of the data for exploration, and then the insight that the user gains may be deeply related to this issue. Thus, the interaction of \textit{retrieving details on demand} is related to the \textit{depth} of the generated insight based on a distinct user intent.

Gotz and Zhou \cite{characterizing} characterized user interaction at four levels of granularity: tasks, sub-tasks, actions, and events. Tasks and sub-tasks are usually domain specific, whereas events, such as mouse clicks, have no meaning without context. In the action level, each action represents a distinct user intent, which makes actions generic across visualization tools yet semantically rich. Interaction categorization based on actions also serves our purpose of using user intent to relate interaction and insight. Furthermore, ElTayeby and Dou \cite{logSurvey} suggested an extra level between actions and sub-tasks that consists of patterns of multiple actions to support the reuse of analysis across different applications.

This work thus explores \textit{which action or pattern relate to which aspect of insight quality} via a case study. First, we redesigned the interaction of MediSyn \cite{medisyn}, a visualization tool that synthesizes five datasets of drug-target relations. MediSyn uses the concept of entities (that is, taking drugs and targets as entities) to afford five types of interactions for insight generation: 1) selecting entities of interest, 2) connecting relevant entities, 3) elaborating by retrieving the details of entity relations, 4) exploring other entities, and 5) entity-based insight sharing. 
We then evaluated MediSyn with 14 participants by asking them to input their tasks, freely explore the data with their tasks in mind, and derive insights by inputting notes. We graded the recorded insights on four aspects, manually extracted seven interaction patterns from the logged interactions, and analyzed the correlations between interactions and insights.
The results show the potential to qualify insights by interactions. Among others findings, exploration actions tend to increase the unexpectedness of insights; the drill-down pattern can lead to insights with high domain values, which resonates with a qualitative analysis. The qualitative analysis of user strategies also uncovered that with domain-specific data, using domain knowledge to guide data exploration helps derive insights with high domain values; users tend to explore public insights when they are stuck with data exploration.
We discuss the implications of this study, lessons learned, and future research opportunities.

\section{Related work}
The traditional task-based evaluation, which measures task time and error, hinders the assessment of the exploratory feature of visualization \cite{toward, rome}. Researchers, therefore, proposed insight-based evaluation instead, which investigates the characteristics of insights derived from visualization tools \cite{biology, longitudinal}. Choe et al. \cite{selfer} and Yang et al. \cite{composing} studied categories of insights derived from various charts, as well as the appropriate charts to use to derive these insights, to gain knowledge on building visualization tools for insight discovery and communication. Smuc et al. \cite{score} evaluated a visualization by analyzing 1) the quantity of generated insights, 2) insight categories, and 3) relations between insights (i.e., how insights build on one another). They found that the third analysis was more informative on improving the design of the visualization tool.

However, analyzing insight alone is often limited. Looking into the insight generation process as a whole can shed more light on improving visualization design \cite{rome}. Mayr et al. \cite{rome} compared three evaluation methods: the task-based method, the insight-based method, and problem-solving strategies. Problem-solving strategies are extracted by analyzing thinking aloud data, eye movement data, and interaction logs. They found that the insight-based approach informs the types of insights that the tool fosters or hinders, whereas analyzing problem-solving strategies helps identify how the tool should be improved. 

The key role that interaction plays in identifying the insight generation, sensemaking, and reasoning processes puts the science of interaction under the focus of the visualization community \cite{science, sensemaking, semantic}. Pohl et al. \cite{interactivity} studied interaction patterns extracted from logs during user visual exploration, and they found that patterns can be valid across visualizations. To support interaction pattern analysis, researchers have proposed several tools. Liu et al. \cite{Sequences} and SensePath \cite{sensepath} used multiple-linked views to support the pattern analysis of user online activities. Liu et al. coordinated multiple levels of granularity of web clickstream data. SensePath captures and displays user actions in temporal order. Analysts can inspect the web page, the recorded video, and the transcribed information of a selected action in other linked views \cite{sensepath}.
Blascheck et al. \cite{VA2, datarich} proposed two visual analytics tools that integrate transcribed thinking aloud data, eye movement data, and interaction logs to assist with the analysis of user studies. These tools support the interactions such as pattern search and comparisons between participants for analysis. Blascheck et al. further used this concept to study the strategies that users employ to discover the interaction techniques available in a visualization \cite{discovery}.

Apart from interaction pattern analysis, other metrics are used to reveal new facets of people's interaction with visualization tools. Two such metrics of exploration uniqueness and exploration pacing were proposed by Feng et al. \cite{pace}. Battle and Heer \cite{tableau} found that users tend to plan and execute a few steps of their interactions at a time as an exploration pace. 

Recorded interaction logs can also be used to infer user tasks \cite{behavior}, to recover reasoning processes \cite{recovering, capturing, helping}, to predict task efficiency and personality traits \cite{waldo}, etc. Dabek and Caban \cite{grammar} built a model that learned from user interactions of solving close-ended tasks and provided suggestions to new users to guide them through the same tasks. Shrinivasan et al. \cite{Connectingthedots} proposed an algorithm that recommends related notes and concepts to users based on their line of analysis. 

Previous work has significantly advanced research on evaluating visualization and understanding users by analyzing user insight or interaction. However, the interrelations between insight and interaction are under-explored. For instance, the thinking aloud data acquired by Blascheck et al. were not further analyzed and connected to insight. There are a few exceptions to this, however. Reda et al. \cite{modeling} proposed a graph that captures transitions between users' mental and interaction states. Each design variation generates a transition graph, which facilitates the discovery of which design variation or interaction type fosters or hinders insight. However, this method omits the effects of interaction sequences on insight generation. 
Guo et al. \cite{casestudy} analyzed the correlations between the types or sequences of interactions and the quantity of insight by asking users to complete open-ended tasks using a visual analytics tool. The results showed, among others, that exploration actions promote the generation of facts, whereas filtering actions inhibit it. With a similar purpose of bridging interaction and insight, this work investigates interaction types and patterns in relation to the quality of insight.

\section{Design considerations}
We propose to use the notion of entities to design the interaction for insight generation, and we review the quality of insight for investigation. 
\subsection{Entity-based interaction design} \label{sec:share}

Entities represent any real-world objects or concepts, such as people, places, and topics.
To design interaction for data visualization, we found the notion of entities useful in supporting visual representation and manipulation, as stated in Klouche et al. \cite{hypercue}. Entities can yield other related entities to support exploration; they can be organized to support pattern recognition; and they can be shared to assist in collaboration \cite{hypercue}. These concepts have been successfully applied to various visualization tools. For instance, Pivotpaths \cite{pivotpath} allows users to pivot on an entity to view its relevant entities in a graph. Jigsaw \cite{jigsaw} coordinates multiple views, and each one shows entity relations from a different perspective to support pattern recognition. Bier et al. \cite{entitycollaboration} proposed an entity workspace that allows users to organize and share their entity graphs and obtain recommendations based on their peers' entity graphs. 

The trend of allowing users to share insight during visual exploration has led to two new design considerations that we identified from previous work. First, the visualization and insight should be linked to support bidirectional exploration \cite{insideinsights, ink}. On the one hand, data exploration should be linked to insight exploration \cite{dynamics, voyager}. Heer et al.'s research \cite{voyager} and InsideInsights \cite{insideinsights} display related insights when users navigate to a visualization to which those insights refer. This can be opportunistic. Heer and Shneiderman \cite{dynamics} later proposed the concept of data-aware annotations, which indicates applying annotations to different views of the same data. We found that entities can support this concept well. The entities that compose a view can change, whereas insights can be dynamically displayed to match the entities under exploration ( e.g., \cite{datasite}).
The other direction is exploring how to use insight to stimulate data exploration.
Bier et al. 
\cite{entitycollaboration} allowed analysts to discover new entities as potential interests from their peers' entity graphs. ManyEyes \cite{manyeyes} and InsideInsights enable restoring the visualization when the user navigates to an insight. This method, however, limits users' ability to derive insight from relating multiple views.

We propose using entities to mediate insight and visualization exploration. For instance, the entities mentioned in the insight can be used to stimulate the visual exploration of their relations. Entities in the visualization can trigger user exploration of related insights.

The second consideration is that mechanisms should be provided to help users rationalize individual insights. Alspaugh et al. \cite{interview} and Madanagopal et al. \cite{practice} identified the key role of analytic provenance in helping data analysts recall, reason, and collaborate through expert interviews. But the provenance feature is not well-supported by existing tools \cite{interview}. 
See Nguyen et al. \cite{survey} and Ragan et al. \cite{characterizingProvenance} for detailed reviews of existing research on visualizing provenance data.
Provenance can be supported by enabling users to attach data sources to notes \cite{annotationgraph, redundancy}, by allowing them to manually create a trail depicting visualization or data flows \cite{graphTrail, explates}, or by automatically capturing and displaying actions \cite{sensepath, analyticTrails, Groth, handoff} or visualization states \cite{vistrail, model, graphical, script}. 

To support insight communication, some research has featured the manual construction of a knowledge graph. A knowledge graph allows users to take recorded visualization states \cite{sensemap}, to create notes \cite{supporting}, or to do both \cite{scalable} as entities and link the entities as causal relations to communicate findings. The manual construction of a knowledge graph may allow users to flexibly map the graphs to their mental models, but this approach requires extra user effort. To provide the rationale of an insight, a trail that leads to the insight and that records semantically meaningful actions as well as their resulted visualization states is attached to the insight in our design.

\subsection{Quality of insight} \label{sec:chara}
To evaluate insights, some research (e.g., \cite{score, selfer, casestudy}) has classified insights by analyzing the content of collected insights. For instance, Guo et al. \cite{casestudy} classified insights as facts, hypotheses, and generalizations. To assess the quality of insight, however, we found that it was important to look at one insight from multiple perspectives rather than simply categorizing it. North et al. \cite{toward} characterized insight as complex, deep, qualitative, unexpected, and relevant. This is for defining insight rather than for quantifying insight for practical use.  
To support the practical and in-depth analysis of insight quality, Saraiya et al. \cite{biology} characterized insight by directness versus unexpectedness, correctness, breadth versus depth, and domain values based on realistic case studies. We use this characterization to qualify insight in this paper, although previous studies that have analyzed these characteristics have not produced promising results. For example, in quantifying the domain value of individual insights, North et al. \cite{comparison} found that only the ``summary" type of insights was more valuable than the others. In Saraiya et al.'s study, most insights were the ``breadth" type rather than the ``depth" type. We think that this may be due to the lack of interaction types supported by the visualizations they studied. The visualizations that North et al. evaluated were mostly static, whereas the ones that Saraiya et al. studied supported mostly standard charts, such as scatter plots and histograms, which facilitated pattern recognition, but limited the user intent of drilling down into the details of individual data items. The limited support of having users elaborate on the data may cause a lack of depth in insights. 
This work would be the first attempt to link multiple types of interactions to these aspects of insight quality. 

\section{The Entity-Based Visualization Tool}
The tool we used synthesizes five publicly available, manually curated drug-target datasets to assist in the research of personalized cancer therapy. \textit{Target} here means point mutations in specific tumors. 

\begin{figure*}
  	\includegraphics[width=0.99\textwidth]{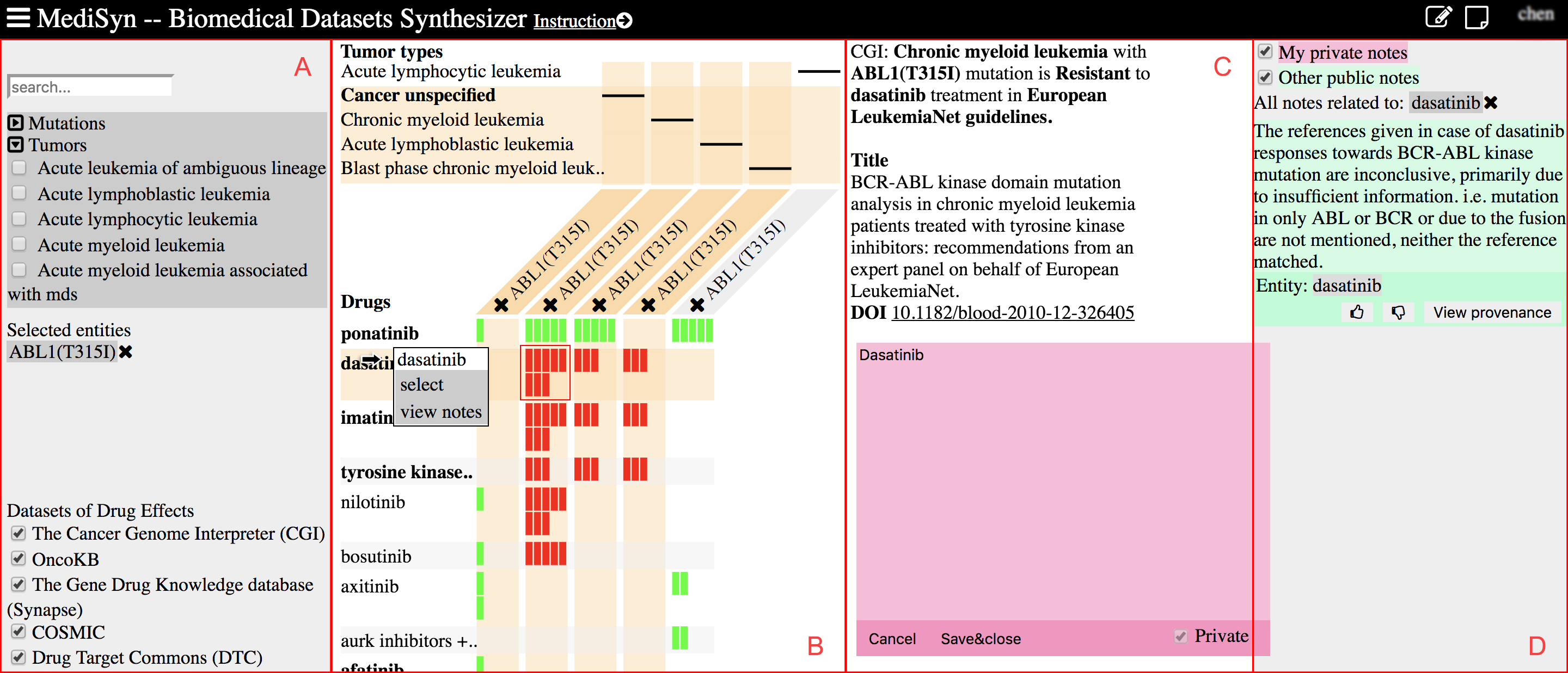}
	\centering
    \vspace{-3mm}
    \caption{The interface of MediSyn consists of (A) the datasets and entities for selection, (B) the visualization, (C) details related to a clicked colored bar, and (D) public and private notes. The pink sticky paper allows users to enter and drag around a note.}
    \label{fig:interface}
\end{figure*}
\subsection{Visualization design}
The visual encodings of data variables have been rationalized in previous work \cite{medisyn}.
MediSyn uses a matrix-based view to relate drugs, mutations, and tumors. As Fig. \ref{fig:interface}B shows, the upper rows depict tumors, the columns denote mutations, the lower rows represent drugs, and the table cells depict their relations. The horizontal line of a cell in the upper rows denotes the tumor to which the drug-target effect refers. In the lower rows, we use colored horizontal bars to depict the corresponding effects and their evidence levels. The hues represent effects. For example, the effects of responsiveness and resistance are depicted by green and red hues, respectively. The length of the bars denotes the evidence level of an effect. As validated in the previous study, MediSyn makes knowledge about genomically informed therapy accessible and evaluable to biologists \cite{medisyn}. Based on the previous study \cite{medisyn}, we had three updates in this version:
\begin{itemize}[noitemsep, leftmargin=0.5cm]
    \item We use discreet bar segments to ease the previous version's perceptual difficulty of comparing the lengths of the bars across columns. From one to five segments, the bars represent ascending evidence levels from preclinical assays to guidelines, such as Food and Drug Administration guidelines.
    \item We extended the number of datasets from two to five. The five datasets result in 427 types of drugs or drug combinations, 827 kinds of mutations, 58 kinds of cancers, and 20,887 total relations. The various datasets contain 19,299, 365, 320, 408, and 495 numbers of relations. 
    \item We use juxtaposed bars to depict values from various datasets to support comparison. The previous design, which used overlaid layers to depict data values from two datasets, is not scalable to the increased number of datasets. In this design, for instance, two juxtaposed bars of the outlined table cell in Fig. \ref{fig:interface}B depict resistant effects from two datasets with varying evidence levels.
\end{itemize}

\subsection{Entity-based interaction design}
To support insight generation, we redesigned MediSyn's interaction. We present the types of interactions based on Yi et al.'s taxonomy \cite{taxonomy}. This taxonomy categorizes interaction based on user intent, which maps to the ``action" tier according to Gotz et al. \cite{characterizing} and serves our purpose of relating interaction and insight by user intent. Additionally, we created a new type of action called ``share" that indicates the user intent of exploring shared insights.
We also discuss how each type of action can support insight generation.

\subsubsection{Select entities of interest}
Considering drugs, mutations, and tumors as entities allows users to explore the data from various aspects.
As Figure \ref{fig:interface}A shows, users can select drugs, mutations, and tumors from the list to add to the view. Once a user selects an entity, its relevant entities across all datasets will be retrieved and depicted in the view for analysis. The user can deselect an entity by clicking the cross icon next to an entity (Fig. \ref{fig:interface}).

\subsubsection{Connect related entities}
Viewing data from various perspectives facilitates insight \cite{understanding, science}. MediSyn allows users to center on an entity to see its connections to other entities in two ways. 
One is to highlight related entities by mousing over the entity, and the other is to aggregate the related entities through the reordering of rows and columns by clicking on the entity, which we call a ``pivot" interaction.
To be specific, when a user mouses over a drug, the columns of its relevant mutations and the upper rows of its relevant tumors are highlighted in a semitransparent orange color. When a user clicks on the drug, its relevant mutations gather to the left of the columns, which pushes the irrelevant columns to the right in animated transitions. The relevant tumors gather to the bottom of the upper rows in the same manner (Fig. \ref{fig:interface}B). The same actions also apply to the other types of entities, mutations and tumors.
For the display of mutations, the same mutation can appear in multiple columns because the same mutation can appear in various tumors. When a user clicks on a mutation, it aggregates the columns of that mutation as well as its relevant drugs and tumors. 

This allows the user to analyze the data from various perspectives, centering on the drug, tumor, or mutation. For example, users can focus on a drug by highlighting or aggregating relevant mutations and tumors, and they can investigate how the effects change across mutations and tumors. The users can also sequentially highlight two tumors and check how the drugs overlap.

\subsubsection{Elaborate by retrieving the details of entity relations}
Being able to view the data in multiple levels of granularity allows users to focus on what is important at the current moment and omit distracting information \cite{entitycollaboration}. With MediSyn, users can retrieve details on demand for further inspection. When a user hovers over a colored bar, the corresponding drug, mutation, and tumor are highlighted, which indicates the intent of viewing an entity relation.
When users click on a colored bar, they can see its details on the right side of the matrix (Fig. \ref{fig:interface}C), which includes the natural language description of the relation and the titles and digital object identifiers (DOIs) of the publication sources. Users can click on DOI links to further review the publications in a new window.

\subsubsection{Explore other entities}
To generate insight, users first need to collect sufficient information \cite{sensemaking}. Exploring other entities is an effective way to do so. 
In MediSyn, exploring other entities is indicated by the user adding an entity to the view while attending to other information. First, in the current view, the user can add an existing entity to the view by mousing over the entity and choosing to ``select" the entity through the drop-down menu (Fig. \ref{fig:interface}B). Afterward, all relevant entities will be added to the view. Second, the user can select the entity to add to the view from an insight display, as articulated next. 

\subsubsection{Entity-based insight sharing}
The user can input an insight by clicking the upper-left paper-and-pen icon (Fig. \ref{fig:interface}). A text area resembling a sticky paper will appear. Afforded by sticky papers, the user can drag and drop the area freely on the interface and can continue to analyze the data while taking notes. By clicking the paper icon next to the note-taking icon, the user can check publicly available notes as well as one's own notes, as shown in Fig. \ref{fig:interface}D. The user can choose to like or dislike notes by clicking the thumbs-up or thumbs-down icon, and view the provenance trail by clicking the ``view provenance" button (Fig. \ref{fig:interface}D).

To support the bidirectional exploration of the insight and visualization we discussed in Section \ref{sec:share}, 
entities mentioned in a note are automatically extracted as tags of the note and are displayed below the note itself, as shown in Fig. \ref{fig:interface}D. The user can explore the entity mentioned in the note by clicking the tag, which results in a select action. On the other hand, visual cues in the visualization imply opportunities to explore related notes without being distracted \cite{entitycollaboration}.
If an entity in the view has related notes available, it will be shown in bold (Fig. \ref{fig:interface}B). As the user hovers over the bold entity, a drop-down menu allows him or her to choose to ``view notes" on part D of Fig. \ref{fig:interface}. 

To help users rationalize others' insights, each insight has a trail attached that records semantically meaningful actions leading to the insight. To construct the visualization of interaction trails, we take a description of an action together with its resulting view as a node and connect them sequentially (e.g., Appended Fig. \ref{fig:prove}). For simplicity of display, the trail will always take the action of changing the numbers of entities in the view as a starting point, such as selecting, de-selecting, and selecting an entity from a note. Sequential select and de-select actions are combined into one node. The interaction of writing the corresponding note completes the trail. In addition, hovering interaction is omitted. See Supplemental Video 1 for a video demonstration of the interaction of MediSyn. Additionally, MediSyn is accessible with a user registration at \url{https://d4health.hiit.fi}. 

\subsection{Use case}
To illustrate the use of MediSyn, we describe a use case collected from the user study we will present later. 
To explore approved drugs for lung cancer, the biologist selected the entity ``lung" from the left list. The selected entity and its related mutations and drugs were then displayed in the visualization. The biologist viewed the details of some relations by clicking on or hovering over colored relation bars sequentially. The relations she checked included that the lung cancer carrying mutation EGFR(T790M) was resistant to erlotinib or afatinib treatment, the lung cancer carrying BRAF(V600E) was responsive to the drug combination of ``dabrafenib + trametinib" based on guidelines, and so forth. She then retrieved all available notes related to the entity ``lung" cancer. The first note in the list was ``Lung cancer carrying EGFR (T790M) are highly resistant to EGFR inhibitors like afatinib and erlotinib. However, combination of afatinib together with cetuximab or nimotuzumab is effective." 

The biologist continued to explore the details of several relations and highlighted three entities---``dabrafenib + trametinib," ``egfr tkis + mek inhibitors in egfr mutant," and ``egfr tkis"---to see their relations to targets. Next, she started to input the following note: ``There currently are only a few approved targeted therapies for lung cancer, such as ceritinib, afatinib and erlotinib, but several more are undergoing clinical trials." During this inputting, she retrieved all notes related to ``ceritinib." Afterward, she moused over the relation that showed that the lung cancer carrying BRAF(G469A) is resistant to egfr tkis treatment based on a case report. She then explored the existing targets of egfr tkis treatment by clicking on the entity. After clicking on a series of entities of drug combinations, including ``egfr tkis + mek inhibitors in egfr mutant," she continued to write, ``Depending on a mutation, different drugs show desired potency. In particular, lung cancer carrying BRAF(G469A) mutation is resistant to egfr tkis treatment but shows response to a combination of egfr tkis and mek inhibitors in preclinical studies." She then saved and closed the note.

\begin{table*}
\centering
\begin{tabularx}{\textwidth}{ >{\raggedright}p{.1\linewidth} | p{.85\linewidth} } 
\hline
 \textbf{Action} & \textbf{Application-specific instantiations} \\
\hline
 Select &  Select an entity of interest from the left list. \\
\hline
Connect & Highlight entities related to an entity by mousing over the entity. \newline Aggregate entities related to an entity by clicking on the entity. \\
\hline
Elaborate & Highlight a table cell depicting entity relations and its related entities by mousing over the cell. \newline Retrieve the details of an entity relation by clicking on a colored bar. \newline Retrieve a publication source by clicking a DOI link. \\
\hline
Explore & Retrieve other entities related to an existing entity by mousing over the entity and clicking ``select" from a drop-down menu. \newline Explore an entity mentioned in a note by selecting the entity from the note display. \\
\hline
Share & Retrieve all available notes. \newline Retrieve notes related to an entity by mousing over the entity and clicking ``view notes" from a drop-down menu. \newline View the provenance view of a note. \newline Like or dislike a note. \\
\end{tabularx}
\caption{Actions and their application-specific instantiations.}
\label{tab:action}
\end{table*}
\section{User study}
To investigate the research question---\textit{which action or pattern relate to which aspect of insight quality}---and explore user strategies of deriving insight using MediSyn, we conducted a lab study.

\subsection{Participants}
In total, we recruited 18 paid participants from a research institute related to personalized cancer therapy. Two were recruited for a pilot study whose research topics were machine prediction of drug-target interactions.
The remaining 16 were recruited for the lab study, but two quit in the middle of the experiment, as one was interested in mutation frequencies in cancer, not drug-target interactions, and the other, a junior researcher, stated she was not able to write notes. The remaining 14 participants ( mean age: 31.36; SD: 5.11; female: 5) had sufficient knowledge to analyze the data. Among them, eight were doctoral students, four were postdoctoral researchers, one was a master's student, and the last was a senior researcher. Their specific focus areas differed. For instance, one focused on lung cancer, another specialized in ovarian cancer, two had expertise in leukemia, and one was studying multiple melanoma and leukemia. Others focused on drug discovery or translational medicine research and did not explicitly point out diseases of focus. 

\subsection{Pilot study}
The purpose of the pilot study was to collect public notes and to have pilot participants go through the procedure of the actual study.
The two pilot participants followed the same procedure as in the lab study, as we describe next. However, their task differed from that of the lab study. They were encouraged to write as many notes as they could and to post them publicly. If they reached a certain quantity of qualified notes, they could get higher amounts of compensation. A qualified note was considered a useful observation on the data and had to contain at least one entity name. As the participants were instructed, the published notes were anonymous to other users to avoid the influence of user identity on the experiment. The pilot study was conducted in the participants' own offices with use of their own computers.

We collected 40 public notes from the pilot study. Another domain expert, one of the authors, screened through these notes and selected 27 as public notes for the lab study. The remaining were screened out because they were either simple observations of drug-target relations for the visualization or purely based on domain knowledge and disconnected from the visualized data. Accordingly, we updated the description of a qualified note for the lab study, which we present next. On average, the public notes contained 3.74 different entities each (SD: 2.26). 

\subsection{Lab study}
The lab study started with a training session. Participants first walked through the functionalities of MediSyn by following a step-by-step interactive tutorial we created using Intro.js \cite{intro}. In total, the tutorial consisted of 12 interactive steps. In each step, the participants carried out 1-2 interactions following the instructions and clicked the ``next" button to move to the next step until they finished the tutorial. After this process, they were instructed to create accounts in MediSyn and to freely explore the tool until they were satisfied. They were encouraged to ask the experimenter questions during the training session. The training session took around 20 minutes. 

In the actual study, the participants were instructed to first input a task by opening a dialog box. Then they could freely explore the interface with the task in mind and write notes. Whenever they changed tasks, they could open the dialog box, input a new task, and continue to explore that new task. We provided some open-ended task examples to help participants start, such as ``to investigate ** disease" and ``to explore effective drugs for ** tumor." They each could finish the experiment when they thought they had written at least five qualified notes. The requirements for a qualified note were as follows: i) it was a useful observation, such as one's findings, hypotheses, or generalizations, on the underlying data based on one's domain knowledge (a simple observation of a drug and cancer mutation relationship from the visualization, for example, not being a useful note); and ii) it included at least one entity name, such as the drug, tumor type, and/or mutation.

During the experiment, we logged user interactions and recorded the screen and voice of each participant with the participants' consent. We did not establish time constraints for the study. Participants were only able to post private notes, and their likes and dislikes of notes were visible only to themselves, as they were instructed. 

After the experiment, participants filled out a questionnaire regarding their subjective feedback about MediSyn. The entire experiment took around 1-1.5 hours.
The lab study was conducted using a Chrome browser in full-screen state on a 27-inch display with an external mouse and keyboard. The display resolution was 2,560 * 1,440 pixels.

\subsection{Measures}
To investigate the research question, we logged types of interactions, time stamps and entities related to each interaction, and the user inputs of the tasks and notes. Hovering interaction was recorded when data were highlighted for more than three seconds.

We then constructed interaction trails from the logged sequential interaction data. The trails we collected always started with an entity-selection action, either from the list or from the public notes. De-selecting all entities, or sequentially selecting and de-selecting entities to yield a completely different set of selected entities, marked the start of a new trail. In each trail, the last note-writing action indicates the end of the trail, as we are interested in actions that lead to insight.

We also manually extracted interaction patterns from the interaction trails. An interaction pattern here denotes a sequence of actions consisting of at least two types of actions. To extract patterns, we considered not just the sequential relations of actions but also the entities on which the participants operated. For instance, suppose a user were to start from select actions, then retrieve the details of an entity relation, and then connect entities by pivoting on one entity. If entities in the connect action did not appear in the previous elaboration action, then we would have counted two patterns, as select -- elaborate and select -- connect. Otherwise, we would have counted this as one pattern: select -- elaborate -- connect. After extracting all patterns from the trails, we took the patterns performed by at least two participants as candidate patterns 
for further analysis.

\begin{table*}[h]
\centering
\begin{tabularx}{\textwidth}{ >{\raggedright}p{.15\linewidth} | p{.82\linewidth} } 
\hline
 \textbf{Characteristics} & \textbf{Grading criteria} \\
\hline
 Directness versus Unexpectedness & Give 1-2 if the note is related to the task. \newline Give 4-5 if the note goes beyond the task description. \newline Give 3 if the note is in between. \\
\hline
Correctness & Give 1-2 if the note is verified to be wrong through the visualization or Internet. \newline
Give 4-5 if the note is verified to be right through the visualization or Internet. \newline
Give 3 if the correctness is difficult to discern. \\
\hline
Breadth versus Depth & Give 1-2 if the note refers to relations of mutation groups, multiple tumor types, and/or drug groups. \newline Give 4-5 if the note refers to one or more specific relations of drugs and mutations. \newline Give 3 if the note is in between. \\
\hline
Domain value & Give 1-2 if the note is a simple observation from the visualization, such as one drug-target relation or a missing value. \newline Give 1-3 if the note complements missing information with domain knowledge. \newline
Give 3-5 if the note reveals patterns across the visualization. \newline Give 4-5 if the note derives hypotheses from the visualization. \\
\hline
\end{tabularx}
\vspace{.5mm}
\caption{Grading criteria of insight characteristics. }
\label{tab:criteria}
\end{table*}

\subsection{Assessing notes}
We collected 86 notes from the study. Removing 27 notes, 25 of which were data requests (notes such as ``FAM46C mutation is quite frequent in Myeloma. Can't find any information in the database.") and two of which were user interface suggestions, we took the remaining 59 notes as insights.
As stated in Section \ref{sec:chara}, following Saraiya et al. \cite{biology}, we characterized the insight quality by directness versus unexpectedness, correctness, breadth versus depth, and domain value.
Two experts, also authors of this paper, first agreed upon detailed grading criteria (Table \ref{tab:criteria}) and then graded the notes on the four aspects from 1 to 5 independently. For instance, directness versus unexpectedness indicates a spectrum (1 to 5) of one insight quality, where grade one indicates direct insights, and grade five denotes unexpected insights. Directness versus unexpectedness was assessed by the relation of an insight to the user-inputted task. Direct insights related to user tasks, whereas unexpected insights were exploratory and went beyond user tasks. Note that a task could have zero or multiple notes. We assigned a note to the last inputted task, as participants were instructed to do.

Correctness indicates whether the insight can be verified or not. If the insight was a hypothesis, we assessed the rationale for the derivation of the hypothesis. Breadth versus depth relates to the scope of the insight. Broad insights refer to multiple entities or entity groups, whereas deep insights relate to a specific aspect of an entity or entity relation. A domain value ranges from low for simple observations to high for pattern recognition and hypothesis derivation.

\begin{figure}[h]
  	\includegraphics[width=0.5\textwidth]{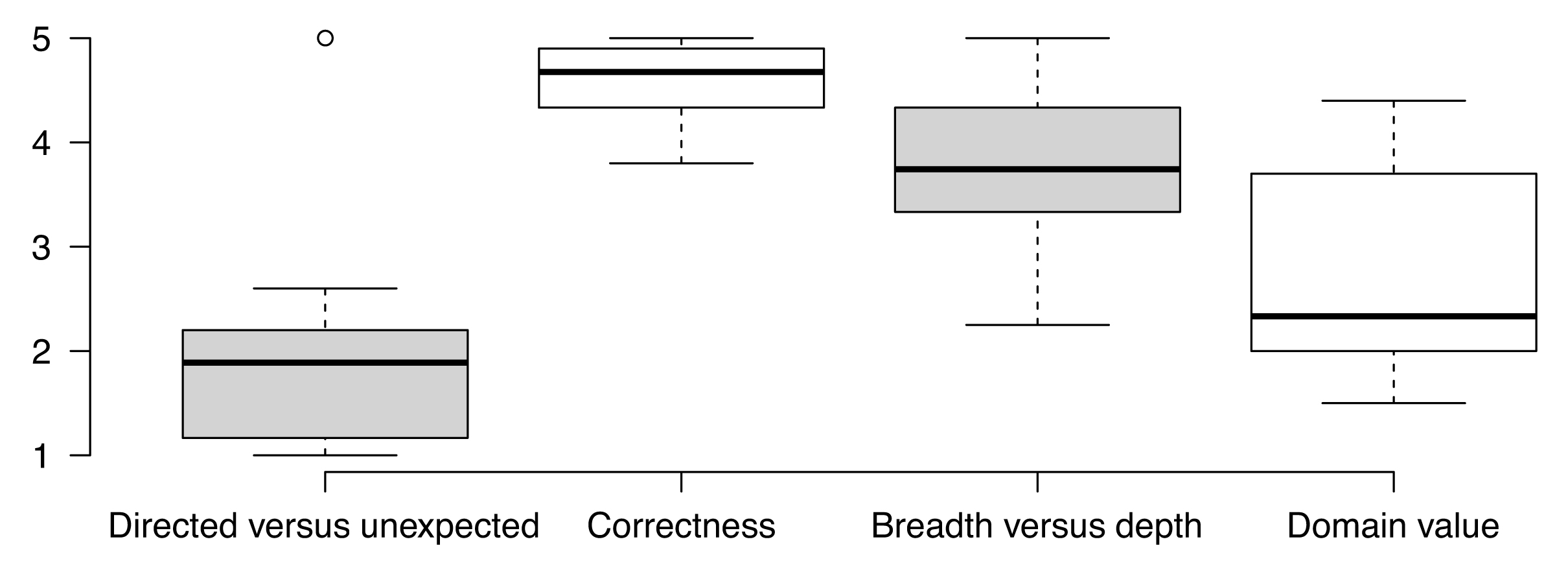}
    \vspace{-5mm}
	\centering
    \caption{The distribution of grades of individual characteristics averaged across participants.}
    \label{fig:chara}
\end{figure}

Kendall's tau-b association test shows a correlation between two raters (p $<$ 0.01). The tau-b values for the characteristics grading of directness versus unexpectedness, correctness, breadth versus depth, and domain value are 0.415, 0.290, 0.423, and 0.245, respectively. Then, for grades differing by more than one point, two raters had a discussion to either change the score or keep their original values. Finally, we took the average of the grades from two raters as the final grades. The distribution of grades of individual characteristics averaged across participants is shown in Fig. \ref{fig:chara}. The insights we collected tended to be direct rather than unexpected, which we discuss in Section \ref{sec:evaluation}. 

We computed the correlations between each pair of characteristics. Two moderate-to-strong correlations existed between the domain value and correctness of insights (r = -0.573, p = 0.032), and the domain value and breadth versus depth of insights (r = -0.607, p = 0.021). The correlations were not statistically significant after the Bonferroni p-value adjustment. Considering all the insights we collected were correct, we could skip the first correlation. However, the second correlation hinted that broader insights tended to have higher domain values in our assessment. 

\begin{table*}[h]
\centering
\begin{tabularx}{\textwidth}{ >{\raggedleft}p{.45\linewidth} | p{.45\linewidth} } 
\hline
  & \textbf{Directness versus Unexpectedness} \\
\hline
\textbf{Exploration action} (Mean: 0.07, SD: 0.10) & $\uptau_b$ = 0.384, p = 0.064, CI [0.0127,  0.7137] \\
\hline
\textbf{Explore -- (connect) -- elaborate} (Mean: 0.09, SD: 0.13) & $\uptau_b$ = 0.489, p = 0.024, CI [0.1800, 0.7574]\\
\hline
\textbf{Explore -- connect} (Mean: 0.15, SD: 0.18) & $\uptau_b$ = 0.375, p = 0.076, CI [-0.0132, 0.6805]\\
\hline
\textbf{Select / Explore -- share} (Mean: 0.11, SD: 0.13) & $\uptau_b$ = 0.326, p = 0.126, CI [-0.0977, 0.7686]\\
\hline
\textbf{Number of interactions} (Mean: 13.57, SD: 11.87) & $\uptau_b$ = 0.363, p = 0.079, CI [-0.0749, 0.8049]\\
\Xhline{3\arrayrulewidth}
 & \textbf{Breadth versus Depth} \\
\hline
\textbf{Elaboration action} (Mean: 0.14, SD: 0.15) & $\uptau_b$ = 0.262, p = 0.202, CI [-0.1707,  0.6110] \\
\hline
\textbf{Explore -- (connect) -- elaborate} (Mean: 0.09, SD: 0.13) & $\uptau_b$ = 0.462, p = 0.033, CI [0.0679, 0.7923]\\
\Xhline{3\arrayrulewidth}
 & \textbf{Domain value} \\
 \hline
 \textbf{Select -- (connect) -- elaborate} (Mean: 0.20, SD: 0.24) & $\uptau_b$ = 0.365, p = 0.081, CI [-0.1056, 0.7754]\\
 \hline
 \textbf{Number of interactions} (Mean: 13.57, SD: 11.87) & $\uptau_b$ = 0.256, p = 0.207, CI [-0.1950,  0.6387]\\
\hline
\end{tabularx}
\vspace{.5mm}
\caption{The report of the assessment includes Kendall's tau-b, bootstrapped 95\% confidence intervals (CI), and p-values. CIs were computed using the percentile method with 2,000 bootstrap replicates. The p-values should be compared with the Bonferroni-corrected value of 0.003 due to multiple comparisons.}
\label{tab:corr}
\end{table*}

\begin{figure*}
  	\includegraphics[width=0.99\textwidth]{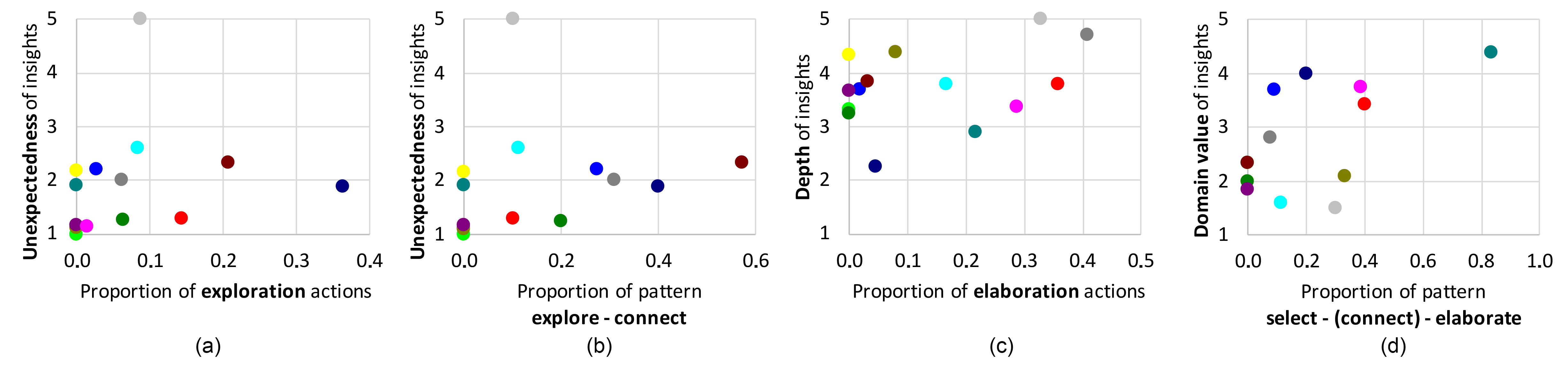}
	\centering
    \vspace{-5mm}
    \caption{Scatter plots show the relations between (a) exploration actions and unexpectedness of insights; (b) pattern of explore -- connect and unexpectedness of insights; (c) elaboration actions and depth of insights; (d) pattern of select -- (connect) -- elaborate and domain values of insights. Participants are differentiated by colors.}
    \label{fig:corr}
\end{figure*}

\subsection{Results}
In total, we collected 37 trails, leading to the 59 insights (Appended Fig. \ref{fig:trail}). Some trails were short, such as ``select, select, and write a note," whereas some others could be as long as 60-70 interactions. Some participants wrote multiple notes in one exploration trail, whereas others wrote one note in each trail. 

\subsubsection{Interaction patterns} \label{sec:pattern}
Interaction patterns indicate higher-level user intents that can be accomplished by carrying out multiple actions \cite{behavior}. The following seven patterns resulted from the study. We named the patterns by the implied user intents. Each action in the pattern can contain arbitrary numbers of the action in succession. 
\begin{itemize}[noitemsep, leftmargin=0.5cm]
\item Drill-down: select -- (connect) -- elaborate. The participants selected an entity or entities from the list, explored entity relations in the view, and then elaborated on certain entity relations. The connect action in this pattern could be zero or more. It was also possible for the participants to carry out the combination of connect -- elaborate in succession multiple times. This pattern featured drilling down into the data of interest. In total, nine users used this pattern.

\item Scanning: select -- connect. The participants selected entities from the list and then explored entity relations through connect actions. Differently from the previous pattern, in this pattern, participants did not dig into the details of entity relations. Thus, the count of this pattern excludes the count of the previous pattern. Eleven participants' trails contained this pattern.

\item Sampling: explore -- (connect) -- elaborate. Compared with the first pattern, rather than merely selecting entities from the list, in this pattern, the participants added an entity or entities to the view through exploration actions, then carried out connect and elaborate actions. Similarly to the first pattern, connect actions existed zero or more times, and the combination of connect -- elaborate actions could recur in succession. This pattern featured exploring and drilling down into the details of other interesting data. Six participants used this pattern. 

\item Probing: explore -- connect. The participants added an entity or entities to the view by selecting from the existing view or the note display and then explored the entity relations through connect actions. Similarly to the scanning pattern, the count of this pattern excludes the previous pattern. This pattern existed in eight participants' trails.

\item Expanding: select -- elaborate -- connect. The participants selected some entities, then elaborated on certain entity relations, and ultimately switched to connect entities in the view. Differently from the first pattern, in this pattern, the participants transitioned from the details of interest to a broader view of data relations. Only three participants used this pattern.

\item Referencing: select / explore -- share. The participants selected or added entities to the view and then explored notes related to certain entities in the view. Seven participants used this pattern.

\item Annotating: select -- write. The participants selected some entities and wrote notes directly without further exploration. Seven participants performed this sequence of actions.
\end{itemize}

Referring to previous work, which also extracted action patterns \cite{casestudy, behavior, interactivity}, we found two common patterns. The drill-down pattern proposed by Gotz and Wen \cite{behavior} features users' intent of narrowing down the analytic focus to a targeted subset of items, which corresponds to the pattern of select -- (connect) -- elaborate in this work and the locating pattern in Guo et al. \cite{casestudy}. The pattern of explore -- (connect) -- elaborate was also discovered by Pohl et al. \cite{interactivity} and Guo et al., the latter of whom led us to name the sampling pattern. 
But for an application, specific actions or patterns can dominate. For instance, a visualization that provides an overview of the data first and allows users to filter the data of interest may not result in the pattern starting with select actions. We thus suggest that the patterns we extracted from the study are specific to a type of applications. 

\subsubsection{Relations between interaction and insight} \label{sec:relation}
To investigate the research question, we took the proportion of each action and pattern from each participant for analysis. The insight characteristics were averaged on each participant. As we considered de-select actions ``passive" reactions to select actions, we did not count them in the analysis.
We omitted too the expanding pattern due to insufficient data, as well as the insight correctness, because it might have been affected by the user interpretation of the visual encodings.
In the analysis, we did not provide hypotheses on the relations of the interaction types and insight characteristics, but rather we investigated each pair to find interesting and noteworthy relations. Depending on whether the values were normally distributed or not via the Shapiro-Wilk test, we calculated Kendall's tau-b or Pearson's correlation to explore their relations. Except for the actions of select and connect, all other actions and patterns were positively skewed. We report all moderate to strong relations in Table \ref{tab:corr}. The p-values should be compared with the Bonferroni-corrected value of 0.003 due to multiple comparisons. 

\textit{Exploring other interesting entities can lead to unexpected insights.}
Among the 14 participants, nine used exploration actions leading to an insight. Only two users explored entities from the note display. Kendall's tau-b test shows a correlation between exploration actions and the unexpectedness of insights ($\uptau_b$ = 0.384, p = 0.064, Fig. \ref{fig:corr}(a)). In addition, two other correlations exist between the unexpectedness of insights and the sampling pattern ($\uptau_b$ = 0.489, p = 0.024), as well as the probing pattern ($\uptau_b$ = 0.375, p = 0.076, Fig. \ref{fig:corr}(b)). 

Ten participants checked other notes through the visualization. Among them, three participants checked the provenance views of notes. We expect that the action of exploring other notes may lead to unexpected insights, but no correlation is found ($\uptau_b$ = 0.148, p = 0.471). However, a moderate correlation exists between the referencing pattern and the unexpectedness of insights ($\uptau_b$ = 0.326, p = 0.126).  

Moreover, a correlation exists between the number of total interactions and the unexpectedness of insights ($\uptau_b$ = 0.363, p = 0.079). It is not so surprising that the more interactions the user carries out, the stronger the tendency is for the insights of that user to go beyond the initial task.

\textit{Select actions have a moderate but not statistically significant correlation to the directness of insights.}
Select actions were used the most (Mean: 0.41, SD: 0.30). We expect that select actions might relate to the directness of insights because with a clear intent in mind, the user might directly select the entities of interest to look up the answer, and consequentially, the answer that he or she gets may be directly related to the task.
\begin{subfigures}
\begin{figure*}[!h]
  	\includegraphics[width=1\textwidth]{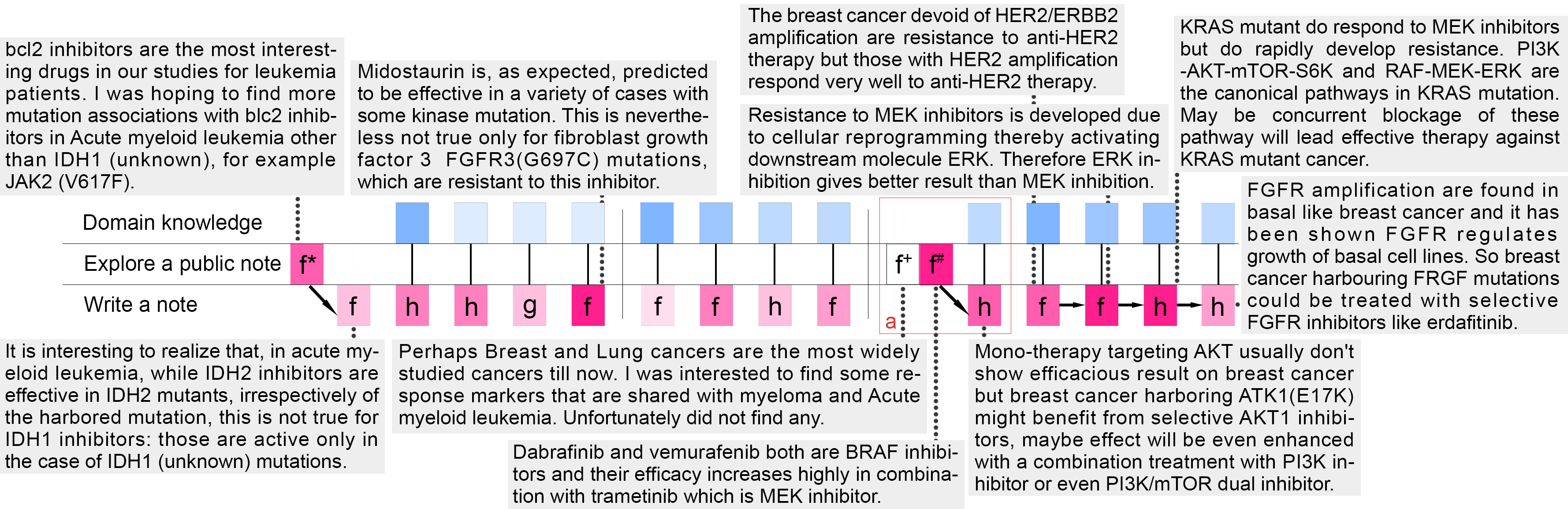}
	\centering
	\vspace{-5mm}
    \caption{This figure and Fig. \ref{fig:referL} demonstrate the insights derived by two groups of participants with the amount of domain knowledge involved and the public notes explored when deriving these insights. This figure shows three participants (participants 12, 9, and 3 in Appended Fig. \ref{fig:trail}) from Group H whose mean domain values of insights are above 3.5. The participants are separated by black vertical lines and listed from left to right in descending order of their indices. A blue node in the first row indicates some domain knowledge is involved in writing the corresponding note where a more highly saturated color means more domain knowledge involved. A purple node in the second or third row indicates a public note explored by the participant or a note written by the participant where a more highly saturated color means more depth, that is, less breadth, of the note. The nodes of each participant are shown from left to right in chronological order. Arrows connect nodes in the same action trail (Appended Fig. \ref{fig:trail}). Dotted lines link several nodes to the actual notes. For symbols on the nodes, `f' denotes a fact, `h' a hypothesis, and `g' a generalization. A node with a special symbol (*, $^+$, or $^\#$) denotes the same node as the one in Fig. \ref{fig:referL} with the corresponding symbol. The red rectangles a is elaborated as case a in Section \ref{sec:quality}.}
    \label{fig:refer}
\end{figure*}
\begin{figure*}[!h]
  	\includegraphics[width=1\textwidth]{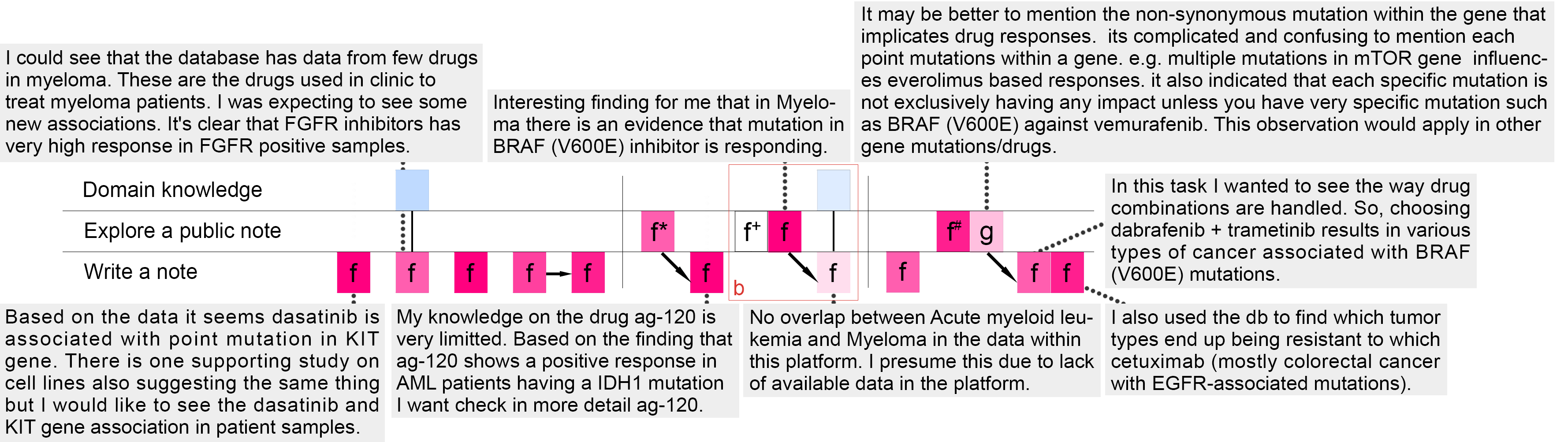}
	\centering
	\vspace{-5mm}
	\caption{The figure applied the same analysis as Fig. \ref{fig:refer} and shows three participants (participants 10, 8 and 7 in Appended Fig. \ref{fig:trail}) from Group L whose mean domain values of insights are below 2.5. The red rectangles b is elaborated as case b in Section \ref{sec:quality}.}
	\label{fig:referL}
\end{figure*}
\label{fig:whole}
\end{subfigures}
We calculated Pearson's r between select actions and the directness versus unexpectedness of insights after removing the outlier in the characteristic of directness versus unexpectedness shown in Fig. \ref{fig:chara}. A moderate correlation exists between select actions and the directness of insights (r(11) = -0.266), but the result is not statistically significant (p = 0.379). Additionally, the annotating pattern has a small to moderate correlation with the directness of insights, and the result is not statistically significant either ($\uptau_b$ = -0.203, p = 0.345).

Ten participants used elaboration actions that led to insights. Most elaboration actions involved retrieving the details of data cells. Only one participant further checked the publication page. Kendall's tau-b shows the existence of a tendency for elaboration actions to lead to deeper insights ($\uptau_b$ = 0.262, p = 0.202, Fig. \ref{fig:corr}(c)). In addition, the sampling pattern shows a positive correlation with the depth of insights ($\uptau_b$ = 0.462, p = 0.033), whereas the drill-down pattern shows no relation to the depth of insights ($\uptau_b$ = -0.035).

Connect actions were used by 13 participants, the second most commonly used type after select actions in deriving insight (Mean: 0.32, SD: 0.17). Its relation to the insights is neutral regarding the characteristic of breadth versus depth (r(12) = -0.047, p = 0.873). 

Regarding the domain values of insights, a positive correlation exists for the drill-down pattern ($\uptau_b$ = 0.365, p = 0.081, Fig. \ref{fig:corr}(d)). This indicates that \textit{the action of drilling down into the details of interest tends to increase the domain values of the derived insights.} Interestingly, but not surprisingly, the annotating pattern has a small to moderate negative correlation to the domain values of insights, although the correlation is not statistically significant ($\uptau_b$ = -0.219, p = 0.315). In addition, a tendency is evident whereby the larger the number of interactions performed, the higher the domain values the insights have ($\uptau_b$ = 0.256, p = 0.207).

\subsubsection{Qualitative analysis of user strategies} \label{sec:quality}
To understand the strategies participants took to gain insights with high domain values, we selected two groups of participants whose average domain values of insights were above 3.5 (Group H) and below 2.5 (Group L), respectively, and analyzed the insights generated by each group. Group H consists of participants 12, 9, and 3 (Appended Fig. \ref{fig:trail}), who took 25.82 (SD:24.91) actions and 8 min 36 sec (SD: 5 min 29 sec) on average to derive an insight. Participants 10, 8, and 7 comprise Group L, and they performed 14.63 (SD 12.24) actions and spent 4 min 37 sec (SD: 2 min 49 sec) on average on deriving an insight. So \textit{the participants who derived high domain value insights used almost twice as many actions and twice as much time as the ones who gained low domain value insights}, which resonates with the statistical analysis. An analysis of actions and patterns show 1) on average, Group L performed 4.6\% sharing actions, 2.1\% more than Group H participants; 2) the drill-down pattern was used by all participants in Group H a total of seven times, whereas only one participant in Group L used this pattern twice, which repeats the finding in Section \ref{sec:relation} that the drill-down pattern can lead to insights with high domain values; 3) the scanning pattern was used four times in each group (by one participant from Group H and all three participants from Group L). Other actions or patterns did not differ much between the two groups. 

Following Smuc et al.'s method \cite{score}, we assessed how much domain knowledge was involved in writing an insight (none, a little, and plenty), and insight categories (fact, hypothesis, and generalization \cite{casestudy}). Two authors of the paper evaluated these two aspects independently and then discussed to resolve any conflicts. Fig. \ref{fig:refer} and Fig. \ref{fig:referL} show the analysis results of Group H and L, respectively. 
We derived three interesting findings by comparing the two figures. First, compared with Group L, participants in Group H were more inclined to involve their domain knowledge during visual exploration. Especially in the first one or two notes, more domain knowledge was involved than in the following notes. The insights they gained were more diverse, as they were not only facts but also hypotheses and generalizations. That indicates \textit{one can generate insights with high domain values when using domain knowledge to guide the visual data exploration}, a similar finding to Smuc et al. \cite{score} that ``an expert's domain knowledge guides the use of a Visual Analytics tool to a great extent." Second, Group H participants started with more broad insights and then went into in-depth discoveries, whereas Group L tended to focus on the details, either initially or throughout the whole exploration process. We cannot deny that users' domain knowledge can guide them to view entities in groups and derive patterns or hypotheses which cannot be easily spotted otherwise. A design suggestion would be to categorize entities, such as grouping mutations by genes, so that patterns are easier to discover.

Third, participants in Group H were more self-motivated and tended to refer to public notes only in the beginning of the exploration or not at all. Based on our collected feedback, participants' opinions on public notes differ. One said that other people's notes were misleading and he preferred to only see his own notes to know what he had explored; another stated that notes were subjective and trusted the data more. A neutral review claimed that the content of the notes was not the same subject as his interests. However, some others had the opposite opinions. One said, ``I wanted to ask the same thing in a public note." Another said that she used notes as a reference. One participant expressed that it was great to see other notes and how others arrived at the conclusions. Regarding user feedback on the provenance view, five users said it helped to explain the logic behind the notes. 
One stated that provenance helps if the selected entities are different from the current user's. That is, the provenance view helped with understanding the scope of an insight. We elaborate on two cases from Fig. \ref{fig:whole} to show how participants leveraged public notes to arrive at their own discoveries.

Case a: The participant input the task ``Breast cancer, the recurrent mutations and effective therapeutics." After some explorations including two actions of retrieving public notes, he said he was confused with what kind of discovery to write. Then he viewed the provenance of two public notes shown in the red rectangle a, Fig. \ref{fig:refer}. Subsequently, he wrote a hypothesis in a similar pattern to the second public note.

Case b: The participant input the task ``investigate Myeloma." Then he selected the entity ``Myeloma," which resulted in five drugs and six mutations in the view. He chose to view notes related to ``Myeloma," which resulted in two public notes shown in the red rectangle b, Fig. \ref{fig:referL}. He clicked the thumbs-up of the first note, the same note as the first public note in case a, and opened its provenance view. He then wrote ``very limited findings on myeloma," which we did not consider to be an insight. Then the participant input another task as ``Investigate the similarities in mutations and drug responses between Acute myeloid leukemia and Myeloma." He selected another entity: ``Acute myeloid leukemia," and wrote a note relevant to the first public note he explored.

The two cases imply that when participants were stuck with data exploration, they would then refer to public notes. In the two cases, public notes helped them to come up with an insight and a new task, respectively.

\subsubsection{Other subjective feedback} 
Three users reported that they gained new knowledge concerning drug sensitivity relationships. Two participants found only the information that they had already known. One participant felt no need for such a tool; the relations between drugs and tumors were enough. 
A lot of the feedback involved requests for more information to be included in the tool. Five participants expressed that more drug-target data were needed. Three participants needed more information on drugs and/or mutations, including mutation frequency in various tumors (two participants) and drug classifications (two participants). Three participants wanted sources of experiments, such as cell lines, mouse models, or patients, along with publications. 

\subsubsection{Learning curve}
To investigate whether usability issues exist which hinder user exploration and insight generation, we assessed the learning curve of MediSyn. Participants tended to take more time and a larger number of actions before writing the first note versus the following notes. The median amount of time that participants spent generating five notes are 460.5, 228.5, 377.5, 181, and 270 seconds, whereas the median steps are 16, 4.5, 8, 3, and 4 steps. Friedman's analysis of variance shows effects among the order of notes considering the time taken ($\chi^{2}$(4) = 12.29, p = 0.015), or the actions taken ($\chi^{2}$(4) = 12.06, p = 0.017). However, the differences are not statistically significant after applying Benjamini-Hochberg corrected post hoc tests. Two participants stated that the tool was difficult to use in the beginning, but they became familiar with it after some time. Five participants expressed that it was easy to view the relations between various drugs and targets with MediSyn, resonating with the fact that most insights are correct from the gradings (Fig. \ref{fig:chara}).

Combining objective and subjective feedback, we conclude that the visual encodings of the data are easy to understand. After the user gained acquaintance with the interaction, which happened around writing the first note, MediSyn had no effects on the efficiency of writing notes.

\section{Lessons learned and future opportunities} \label{sec:discussion}
Though the number of participants is not large in the study, results still support the possibility of qualifying insights by interactions. Exploration actions tend to increase the unexpectedness of insights, whereas the drill-down pattern can lead to insights with high domain values, confirmed by a qualitative analysis as well. However, results also show, using domain-specific data, that the involvement of domain knowledge during the exploration can affect the quality of generated insights. Also, considering that a case study from a single domain may limit the generalizability of the findings, we propose that further research with generic types of interactive visualizations can alleviate the effects of domain knowledge on insights quality and has the potential to positively impact this area. Moreover, personality traits have been shown to influence the number of insights generated by Green and Fisher \cite{personality}. How personality traits affect the quality of insight is an interesting question for future exploration. We discuss our reflections on this study next.

\subsection{Characterizing insights}
We used the four metrics from Saraiya et al. \cite{biology} in this study to qualify insight, although the quality of insight can not be limited to these metrics, such as the one Chris North proposed as simple versus complex \cite{toward}. This study demonstrated that other useful characteristics can also be correlated to various actions through user intent. The interaction logs and insights we collected can be reused. However, assessing the quality of insight can be laborious. To make the assessment process efficient, we found it important to have detailed grading criteria, along with some examples to ensure that the raters were on the same page.

Specific to the visualization we studied, we found that participants frequently spot that specific drug combinations are more effective to certain targets than the individual drugs. This type of insight could be an opportunity for insight automation, which suggests this type of pattern as mixed-initiative analysis \cite{unified}.

\subsection{Interacting with entities}
Abstracting various real-world objects or concepts to entities, on the one hand, simplifies the way of designing interaction. For instance, we design the interaction with one or more entities, rather than designing for drugs, mutations, and tumors separately. On the other hand, similarly to the concept of object-oriented programming, the interaction can be flexibly extended to various numbers of entities, and entity affordances can be applied to any real-world data with similar relations. Thus, abstracting drug-target data to entities implies that MediSyn is not limited to this specific type of data, but rather can be applied to any real-world datasets which represent similar relations. 

From the interaction trails (Appended Fig. \ref{fig:prove}), we recognized that the same actions are prone to appear in chunks, especially for the select and de-select, connect, and elaborate actions. Reflecting back on participants' suggestions of classifying drugs, we suggest that instances of entities can be flexibly grouped based on some criteria to avoid manipulating one by one and instead do so by groups. In our case, mutations can be grouped by genes, and the drugs by approved drug, investigational drug, etc., as one participant proposed. In this way, the users can select entities by groups and look at their relations on a group level to facilitate pattern recognition, as also discussed in Section \ref{sec:quality}.

\subsection{Insight-based evaluation}\label{sec:evaluation}
We have some considerations on evaluating whether a user's open-ended exploration goes beyond his or her original goals, such as supporting serendipitous discovery. In this study, we asked participants to input tasks whenever they came up with new ones. This method may not adequately capture the directness versus unexpectedness of insights. Some participants tried to stick with the task they had input and wrote notes, whereas some others wrote new tasks inspired by exploration. For example, in case b of Section \ref{sec:quality}, the participant was inspired by a note and input a new task, from which we expected the insight generated to be originally recognized as ``going beyond the original task." This caused most insights to be graded as direct (Fig. \ref{fig:chara}). 

On the other hand, input tasks in the middle of data exploration might interrupt the flow of one's thinking. Saraiya et al. \cite{biology} asked participants to identify all tasks before exploration. However, we ponder that during exploration, when participants become familiar with the tool, they may come up with new tasks from their knowledge databases. 
Guo et al. provided an open-ended task to all participants as they studied a generic visualization. This may bring two other challenges. First, there is the question of how open-ended the task should be to allow serendipitous discovery. Second, as proposed by North et al. \cite{comparison} and Gomez et al. \cite{mixed}, participants need to have enough motivation to explore unfamiliar data and generate insight. We need to address these challenges when studying a generic visualization.

But with familiar data, in this study, we found that ``curiosity" is another issue to entice users to go beyond their specific area and generate insight. We observed that some participants tried to focus on their specialized cancer types and did not explore other diseases, so they did not find new information with MediSyn. This phenomenon may also be attributed to the controlled study. An insight-based longitudinal study \cite{longitudinal} in one's natural working environment may reveal other interesting results, which would be worth future exploration.

\section{Implications to knowledge-assisted visualization}
Integrating domain knowledge in visualization, such as the user settings of visualization parameters, to support data exploration, also known as knowledge-assisted visualization \cite{data}, is trending but facing some challenges. This study sheds some light on addressing the challenges in the following ways.

First, this study suggests two possible ways to alleviate the difficulty of evaluating shared insights. Compared with visualizations for general content, domain-specific visualizations have limited numbers of domain experts spreading all over the world. Noises in shared user insights, such as comments without domain values, can be more difficult to discover and can conversely cause more serious consequences. The results of this study show, on the one hand, that visualizing insight provenance can help to explain the logic behind the insight. On the other hand, it is promising to qualify insights through the analysis of their interactions. As an example from the study, the drill-down pattern is shown to be positively correlated to the domain values of insights, whereas the pattern of select -- write tended to be negatively correlated to it. This implies that the visualization tool can possibly predict the level of the domain values of insights by analyzing their interaction patterns.  
The study also showed that one insight characteristic can be related to multiple interaction types or patterns (Table \ref{tab:corr}). Thus, multiple correlations can be evaluated, which may amount to increased confidence in insight characterization.
However, as stated in Section \ref{sec:pattern}, interaction patterns can be specific to certain types of visualizations. Supporting insight characterization in various applications first requires the intervention of experts to learn the relations between insights and application-specific actions, which invites future exploration.

Moreover, qualifying insights by interactions can assist in personalized insight recommendation. 
As the study shows that elaboration actions tend to relate to the depth of insights, MediSyn can recommend in-depth insights to users who elaborate on similar entities, as they are potentially looking for this type of insights. Insight recommendation can be helpful when users encounter limitations with data exploration, as this analysis shows (Section \ref{sec:quality}). 

\section{Conclusions}
This paper presents a case study that investigates \textit{which interaction type or pattern relate to which aspect of insight quality}. First, we used the concept of entities to devise the interaction of a visualization tool---MediSyn---for insight generation. MediSyn synthesizes five drug-target datasets and supports five types of interactions: 1) selecting entities of interest, 2) connecting relevant entities, 3) elaborating by retrieving the details of entity relations, 4) exploring other entities, and 5) entity-based insight sharing. The action of entity-based insight sharing includes 1) using entities to link the visualization and public notes to support the bidirectional exploration of the two components; 2) allowing users to view insight provenance, which records semantically meaningful actions and their resulted visualization states to help rationalize public notes. 

To investigate the research question and user strategies of deriving insight, we conducted a study with 14 participants. During the study, they were instructed to input a task, then freely explore the data with their task in mind and generate insights. We then had two raters grade four aspects, i.e., directness versus unexpectedness, correctness, breadth versus depth, and domain value, of the recorded insights. Apart from the five types of interactions that MediSyn supports, we extracted seven interaction patterns from logged interactions to analyze their relations to aspects of insight quality. Results showed the potential to qualify insights via interactions. Among other findings, exploration actions can lead to unexpected insights; the drill-down pattern positively correlates to the domain values of insights, which is also confirmed by a qualitative analysis. The qualitative analysis of user strategies also reveals that using domain knowledge to guide exploration can increase the domain value of derived insights. To alleviate the effects of domain knowledge on insight quality, we propose that future work on a generic type of visualization can further this research, though we should address user motivation issues by asking them to explore unfamiliar datasets and derive insight. Moreover, how user personality traits relate to the quality of insight is an interesting research question for future exploration.

These results also imply that future research on analyzing interaction types and patterns arriving at an insight can help assess the quality of insight for knowledge-assisted visualization. Opportunities exist for visualization to recommend insights to users who are looking for insights of a certain quality inferred by analyzing their interactions.

\section{Supplemental material}
Supplemental video 1 is a video that provides a demonstration of the five interaction types that MediSyn supports and an overview of the user study. (MP4 67,409 kb)
\acknowledgments{This work is supported by the Academy of Finland (grants no. 295504, and 286440).}

\bibliographystyle{abbrv-doi}

\bibliography{template}

\newcommand{\beginsupplement}{%
        \setcounter{table}{0}
        \renewcommand{\thetable}{A\arabic{table}}%
        \setcounter{figure}{0}
        \renewcommand{\thefigure}{A\arabic{figure}}%
     }

\appendix

\beginsupplement

\begin{figure*}
  	\includegraphics[width=0.34\textwidth]{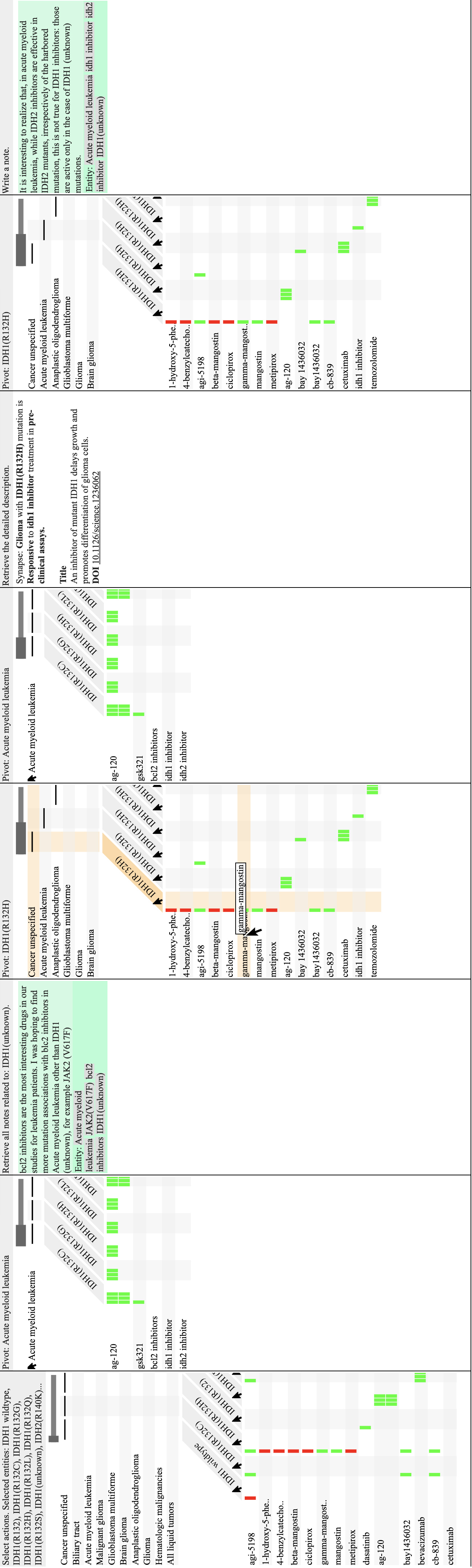}
	\centering
    \caption{An interaction trail that leads to a note. Each action is represented by a description of the action at the top and its result view. Connect actions are supported on the result views. }
    \label{fig:prove}
\end{figure*}

\begin{figure*}
  	\includegraphics[width=0.85\textwidth]{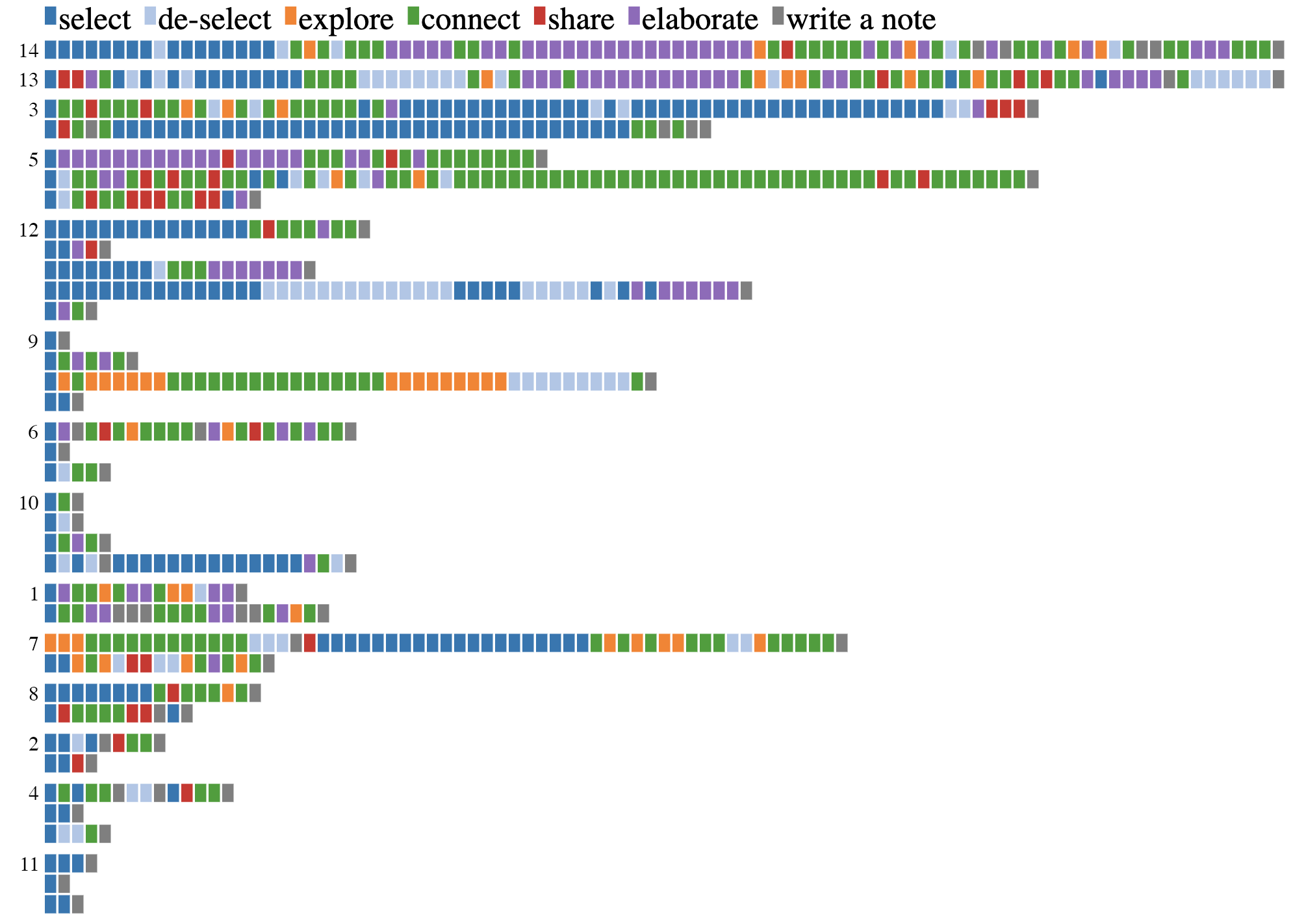}
	\centering
    \caption{Thirty-seven interaction trails in 37 rows. Rows are aggregated by participants and attached with participants' indices. The rows of each participant are ordered chronologically.}
    \label{fig:trail}
\end{figure*}
\end{document}